\documentclass{iopart}
\usepackage{graphicx}

\begin{document}
\title{Veto Studies for LIGO Inspiral Triggers}


\author{Nelson Christensen$^1$ for the LIGO Scientific Collaboration}
\address{$^1$Physics and Astronomy, Carleton College,
Northfield, MN 55057, USA
}

\ead{nchriste@carleton.edu}

\date{\today}


\begin{abstract}
LIGO recently conducted its third scientific data run, S3. Here we summarize the veto and data quality studies conducted by the LIGO Scientific Collaboration in connection with the search for binary inspiral signals in the S3 data. LIGO's interferometer channels and physical environmental monitors were monitored, and events in these channels coincident with inspiral triggers were examined. 
\end{abstract}


\section{Introduction} 
\label{intro}

The Laser Interferometer Gravitational Wave Observatory (LIGO)~\cite{LIGO-DET} is approaching its target sensitivity, and numerous scientific studies by the LIGO Scientific Collaboration (LSC) are in progress. The analysis of the data from the first scientific run, S1, is now complete, and the results of searches for
continuous waves from pulsars~\cite{LIGO-CW}, the ``inspiral'' (orbital decay)
of compact binary systems~\cite{LIGO-IN}, short bursts~\cite{LIGO-BU}, 
and an isotropic stochastic background~\cite{LIGO-ST} have been published. Studies of the data from the second scientific run, S2, are also now being presented~\cite{LIGO-CW-S2,LIGO-BU-S2}. LIGO's third scientific run, S3, is now complete, and the data is currently being analyzed.

The LSC is actively searching for signals from the inspiral of compact binary objects. Descriptions of the matched filtering methods used in this search on the S1 and S2 data have been presented~\cite{LIGO-IN,IN-GWDAW8}. Candidate inspiral events, or {\it triggers}, are produced when the matched filter exceeds a signal-to-noise (SNR) threshold. A $\chi^2$ test is also applied to the events in order to check that the frequency distribution of the signal power is consistent with an expected binary inspiral signal. 

Detector characterization is an important part of the binary inspiral search process. Sections of bad data due to poor interferometer performance or environmental disturbance should be excluded. When we can quantify periods of time where the data is bad, and we can identify the source of the problem, we produce a data quality (DQ) warning, or what we call a DQ flag. When there are numerous inspiral triggers that are in coincidence with short duration glitching in interferometer or environmental channels there is an opportunity to develop a veto. 
The DQ flags and vetoes employed by the LSC in their examination of the S1 and S2 data were described in \cite{S2veto}. In this paper we present a summary of the veto and DQ study results from the S3 data, and state how they will be used in the binary inspiral search.

LIGO's S3 run spanned the time period of 70 days from 31 October, 2003, until 9 January, 2004. The sensitivity of the interferometers was much improved over their S2 sensitivity, and hence the environmental monitors took on increased importance. Acoustic isolation work has also dramatically reduced acoustically generated events. Still, large {\it glitches} would occasionally occur in the gravity wave output channel of the detectors, and these would often be coincident with glitches in interferometer channels, or an environmental disturbance. 

LIGO designated {\it playground} sections of the data, whereby veto studies could be conducted without influencing the statistical validity of the remaining data. For every 6370 seconds, 600 seconds of data are set aside from all three interferometers for the playground. Each segment begins at an integer multiple of 6370 seconds. The playground constitutes 9.42 \% of the total run. A sample begins in each solar hour twice every three days. The veto results presented here come from studies on the S3 playground data.

The rest of the paper is organized as follows. In Section~\ref{DQ} we summarize the data quality flags that we have found to be useful with the S3 data. Section~\ref{S3V} contains a description of specific vetoes which we are considering for implementation in the final search for inspiral signals in the S3 data. A summary is given in Section~\ref{Disc}.  In the course of this paper we refer to the 
4 km interferometer at Livingston, Louisiana, as L1, and the 4 km and 2 km interferometers at Hanford, Washington, as H1 and H2 respectively.

\section{Data Quality Checks for the S3 Analysis}
\label{DQ}

The S3 data from LIGO was monitored in a number of different ways. Various problems caused data to be excluded: missing calibration lines, an unlocked interferometer, data acquisition overflows, and invalid timing.

When we develop vetoes we try to implement a scheme whereby bad sections of data are efficiently eliminated. The inspiral search triggers used in our veto study had a threshold of SNR $>$ 6, but no $\chi^2$ cut. A good veto would have a large use percentage (percentage of veto triggers that veto at least one inspiral event), a large veto efficiency (percentage of inspiral triggers eliminated), and a small deadtime (percentage of science-data time when veto is on). The veto channel needs to be appropriately filtered, and an event size threshold set. The time window about a veto trigger is another parameter in the veto study. The filter frequency band, the threshold, and the length of time about the veto trigger will all affect the use percentage, veto efficiency and the deadtime.

An interesting and new DQ flag implemented in the S3 study is the monitoring of elevated acoustic signals at the interferometers recorded with microphones~\cite{S3-burst-veto}. A number of these events were determined to be airplanes flying over the interferometer, with the sound then coupling to the instrument through the ground. During the S3 H1 playground data, 9 out of 13 of these elevated acoustic events produced H1 inspiral triggers, yielding a use percentage for this veto  of 69\%.  A veto window of order 60s will be used, and since relatively few airplanes contaminate the data the resulting deadtime will be negligible (about 1 \%), as will the veto efficiency, 0.68\%.

Seismic activity was closely monitored throughout the S3 period. Times with elevated seismic activity were seen to cause glitches that were identified as inspiral triggers. Two effective seismic DQ flags were generated for use on the data from the two LIGO Hanford interferometers. The seismic events were observed with a seismometer, H0:LSC-LVEA\_SEISZ, at the central building at the LIGO Hanford Observatory (LHO). 

\begin{figure}[tb]
  \begin{center}
    \includegraphics[width=5.0in,angle=0]{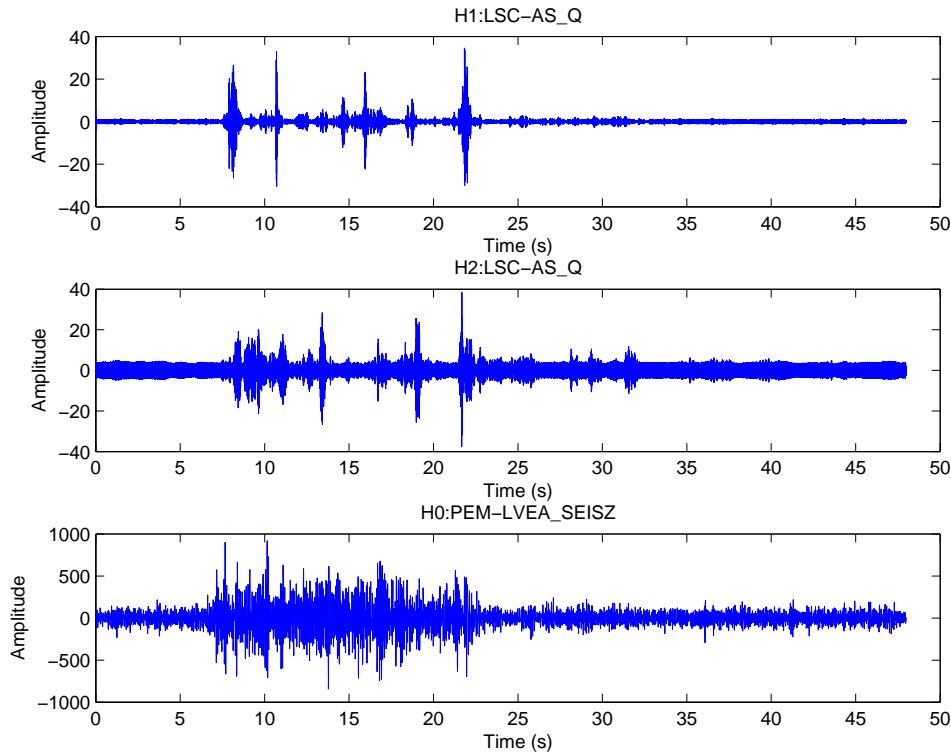}
  \end{center}
\caption{An example of a seismic event at LHO that produced simultaneous {\it glitches} in the gravity wave output channels for H1 and H2. The AS\_Q data displayed here has passed through a 100 Hz to 1 kHz bandpass filter, while the seismometer data displayed in this figure was filtered with a 10 Hz to 20 Hz bandpass.} \label{fig1}
\end{figure}

One class of seismic events were characterized using a program called {\it
glitchMon}~\cite{glitchMon}.  The data was band-passed filtered (2 to 20 Hz), and events of magnitude $9\sigma$ and larger were designated as vetoes. Subsequent to S3 it was determined that these glitches were generated by a liquid nitrogen dewar responding to diurnal variations in temperature. This dewar has now been insulated and the noise generation eliminated. This DQ flag was very important in that it eliminated coincident H1 and H2 triggers, as is the case for the event in Fig.~\ref{fig1}. Often these events caused the interferometers to lose lock, but for those times when the interferometers remained locked this veto had a 100\% use percentage for both H1 and H2. We will set a 20s DQ flag window around the identified seismic trigger. Because of the infrequency of these seismic events the resulting deadtime is insignificant (0.02\% for H1).

The other seismic DQ flag for LHO used data from the same seismometer, H0:PEM-LVEA\_SEISZ. For the generation of the DQ flag the data was band-pass filtered from 3 to 10 Hz. The root mean square (RMS) of the data was calculated, and periods when the RMS was large were excluded. It is believed that this veto is effective in identifying periods when gravel trucks were driving in the vicinity of LHO, plus other seismic events. Fig.~\ref{fig1} shows a time series of one of these seismic events, along with the data from the gravity wave output channels of the H1 (H1:LSC-AS\_Q) and H2 (H2:LSC-AS\_Q) interferometers; this signal is consistent with what is observed due to passing gravel trucks. This DQ flag will exclude about 1.5\% of the S3 H1 data, and its use percentage is 53\%.  

It has been observed that there is an increase in {\it glitches} in the AS\_Q channels of the interferometers for some hours after people have entered the enclosures housing the dark output ports of the interferometers for service work. This increase in glitches is produced by increased numbers of dust particles passing through the dark port beam. The LSC will exclude periods of data with large dust activity from its inspiral upper limit studies, but will analyze the data in a search for inspiral events. The dust flag, for example, would produce a relatively large deadtime for H1 of 4.2\% (with a veto efficiency for H1 inspiral triggers of 9.9\%).

\begin{figure}[tb]
  \begin{center}
    \includegraphics[width=5.0in,angle=0]{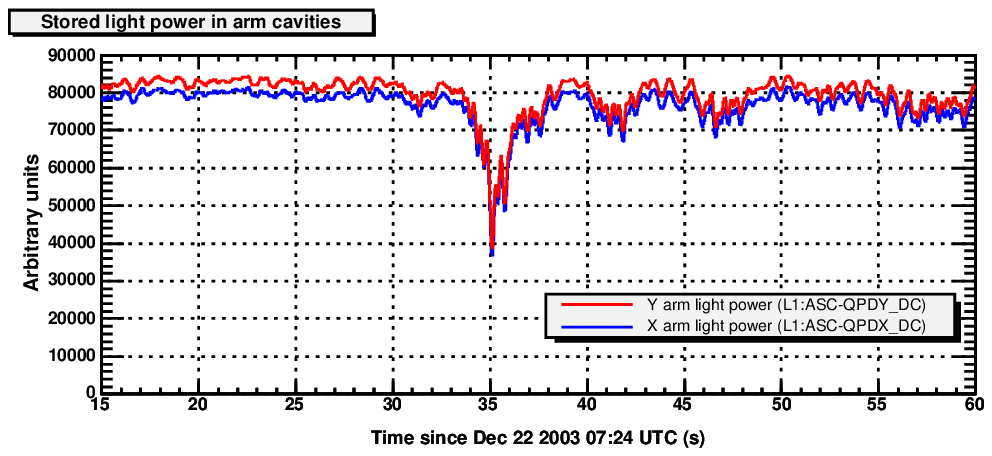}
    \includegraphics[width=5.0in,angle=0]{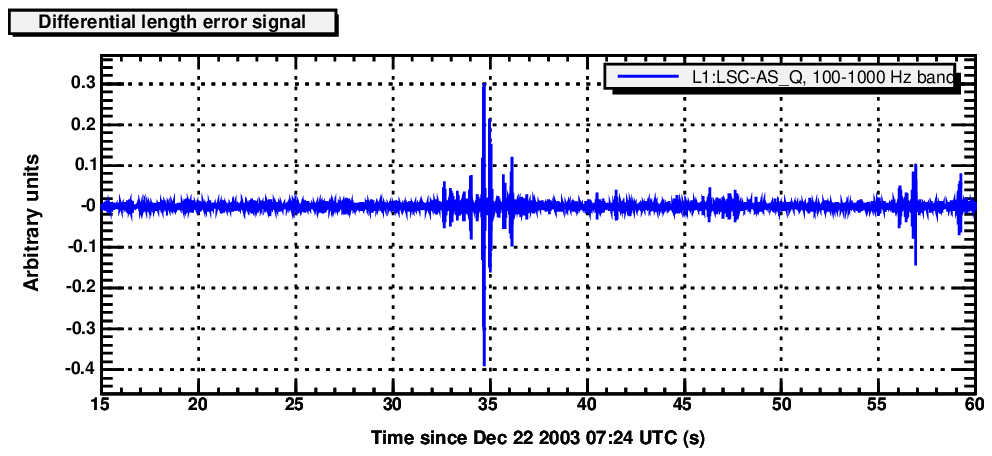}
  \end{center}
\caption{Displayed is a period of time when the light in the Fabry-Perot arms of the L1 interferometer had a dip in intensity. The photodetector signals that monitor these light levels are referred to as L1:ASC-QPDX\_DC and L1:ASC-QPDY\_DC. Also shown is the simultaneous glitch observed in L1:LSC-AS\_Q.} 
\label{fig2}
\end{figure}

Another interesting DQ flag is associated with times when the amount of light stored in the Fabry-Perot arms of the interferometers dips in intensity.
A dip was defined as a decrease in light level, relative to
the average over the previous ten seconds, of at least 15\% for L1 or
5\% for H1 or H2. During the S3 playground period there were nine observed light dips at L1, with four of the times associated with inspiral triggers. Fig.~\ref{fig2} shows an example of one such event. In H1, there were no light dips identified during playground times that were filtered for inspiral triggers. In H2, there were three dips identified during playground times that were filtered, and two of these corresponded to actual inspiral triggers.

\section{S3 Vetoes}
\label{S3V}
Much work went into identifying signals from interferometer channels or environmental monitors that were consistently associated with inspiral triggers. Starting with the inspiral triggers from the S3 playground all of the relevant channels were examined by eye, using numerous different filters. When candidate veto channels were identified veto triggers were generated using glitch finding algorithms. The glitchMon program~\cite{glitchMon}, mentioned above, was used. In addition, {\it KleineWelle} (KW)~\cite{KW}, a wavelet based event finding algorithm was employed. KW was used extensively in the veto studies for LSC burst search group~\cite{S3-burst-veto}. The veto triggers are generated after filtering the data (usually high-passed). Different veto trigger thresholds are tried, and a decision on the usefulness of the veto is based on the veto efficiency, deadtime, and use percentage. 

\begin{figure}[tb]
  \begin{center}
    \includegraphics[width=5.0in,angle=0]{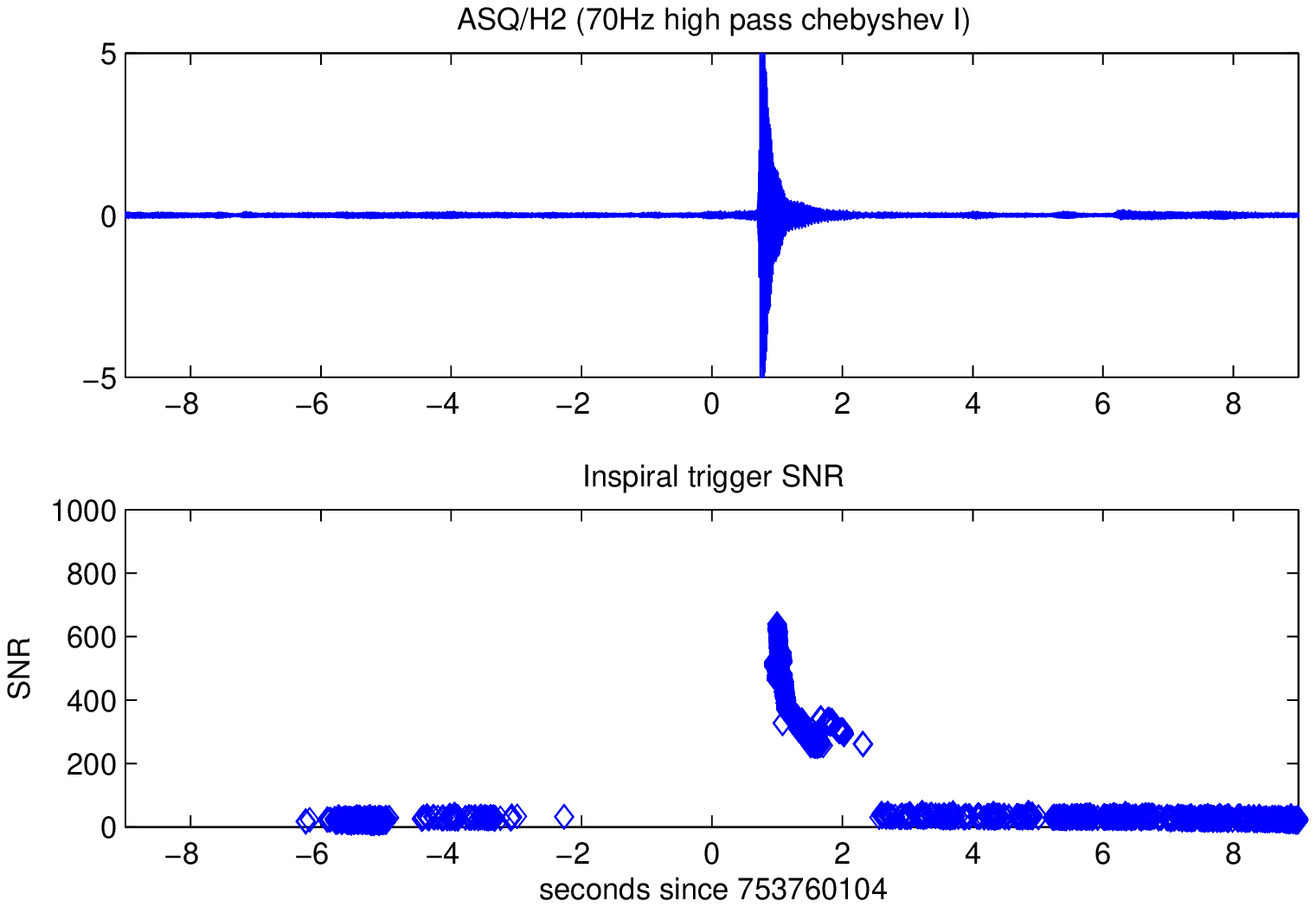}
  \end{center}
\caption{A glitch in H2:LSC-AS\_Q, and the resulting distribution of inspiral triggers.
} \label{fig3}
\end{figure}

The temporal distribution of inspiral triggers produced can be quite complicated due to the nature of the inspiral triggers~\cite{S2veto}. A glitch can cause a large number of inspiral templates to respond, but their reported coalescence times can be quite different. This can be observed with an example glitch from H2 displayed in Fig.~\ref{fig3}. The asymmetric distribution of inspiral triggers about the glitch in H2:LSC-AS\_Q led us to design veto windows that started just before the glitch, and then extended for a few seconds afterward. 

\begin{figure}[tb]
  \begin{center}
    \includegraphics[width=5.0in,angle=0]{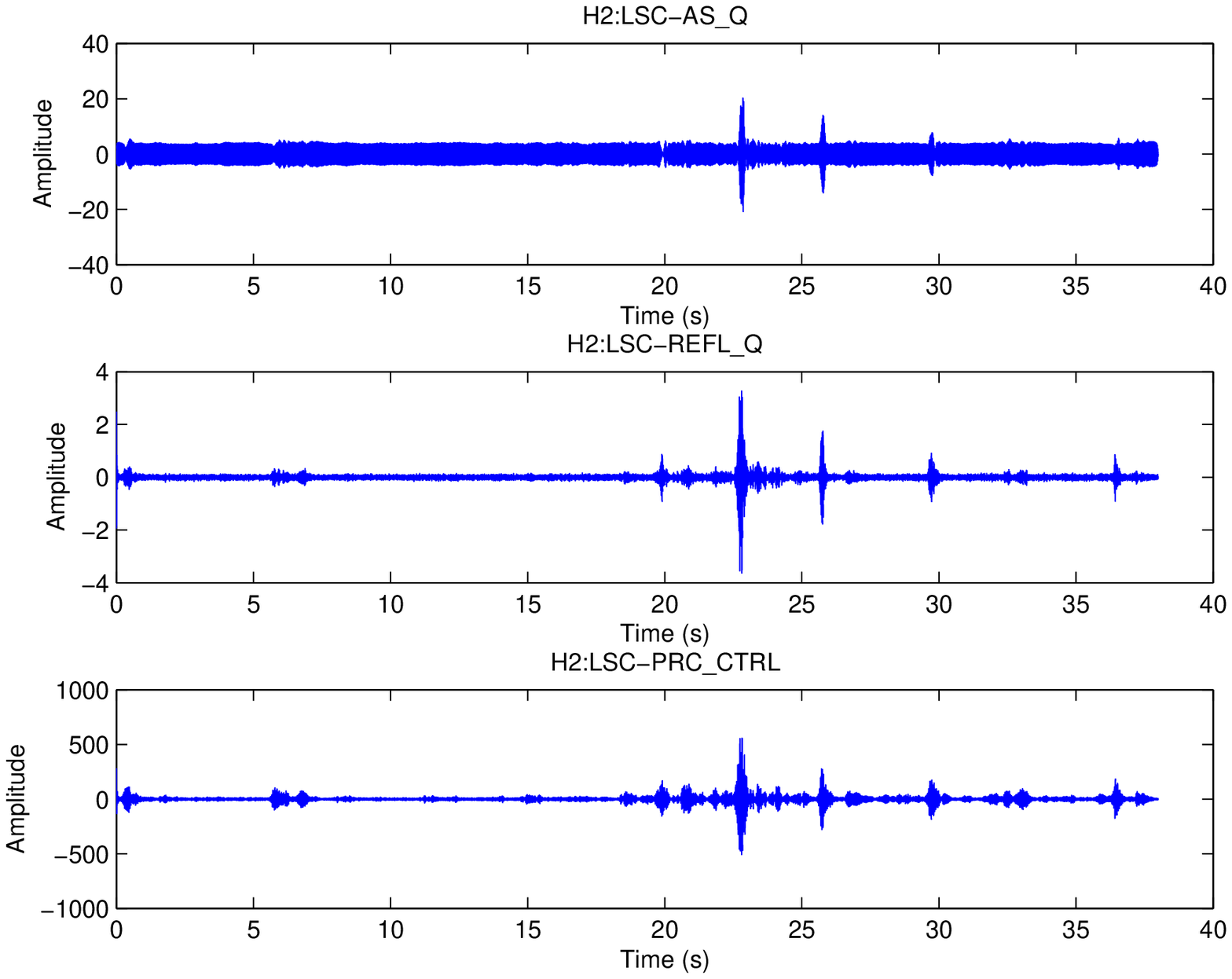}
  \end{center}
\caption{A glitch in H2:LSC-AS\_Q that is also observed in the interferometer signals H2:LSC-PRC\_CTRL and H2:LSC-REFL\_Q. H2:LSC-AS\_Q and H2:LSC-REFL\_Q have been band-passed filtered from 50 Hz to 250 Hz, while H2:LSC-PRC\_CTRL was filtered from 30 Hz to 230 Hz.
} \label{fig4}
\end{figure}

Two good candidate veto channels were found for use with H2 S3 inspiral triggers. H2:LSC-PRC\_CTRL is a control signal ($\propto$~force applied) in the feedback loop that keeps the recycling cavity resonant. H2:LSC-REFL\_Q is an error signal produced by the light coming from the bright port of the interferometer and traveling back toward the laser, and is generated from the motion of the front mirrors of the Fabry-Perot cavity arms or the power recycling mirror. Because each of these channels respond to excitation in the vicinity of the power recyling optics, both of these channels veto similar glitches in H2:LSC-AS\_Q. As an example, see the event displayed in Fig.~\ref{fig4}.

With inspiral vetoes we did not want the deadtime to exceed 0.5\% or so. With that in mind it was found that an effective veto could be created with the channel H2:LSC-REFL\_Q. The data was processed by glitchMon, using a 100 Hz high-pass, and a threshold of $6\sigma$. A veto window was created from 1s before the trigger to 10s after it. With these settings 27\% of all S3 H2 inspiral triggers in the playground data were vetoed, with a deadtime of 0.5\% and a veto use percentage of 71\%. The veto efficiency was 42\% for S3 H2 inspiral triggers with SNR $>10$. 

The S3 H2 veto triggers from channel H2:LSC-PRC\_CTRL were created using the KW program~\cite{S3-burst-veto,KW}. The data were first high-pass filtered at 70~Hz and then whitened. The resulting KW triggers had a threshold {\it significance}~\cite{S3-burst-veto,KW} of 1600; the significance is a function of
the probability of randomly detecting an event with greater signal energy
in perfect white noise.
The veto window again extended from 1s before the trigger to 10s after it. With these settings 26\% of all S3 H2 inspiral triggers in the playground data were vetoed, with a deadtime of 0.6\% and a veto use percentage of 66\%. The veto efficiency was 40\% for S3 H2 inspiral triggers with SNR~$>10$. 

\begin{figure}[tb]
  \begin{center}
    \includegraphics[width=5.0in,angle=0]{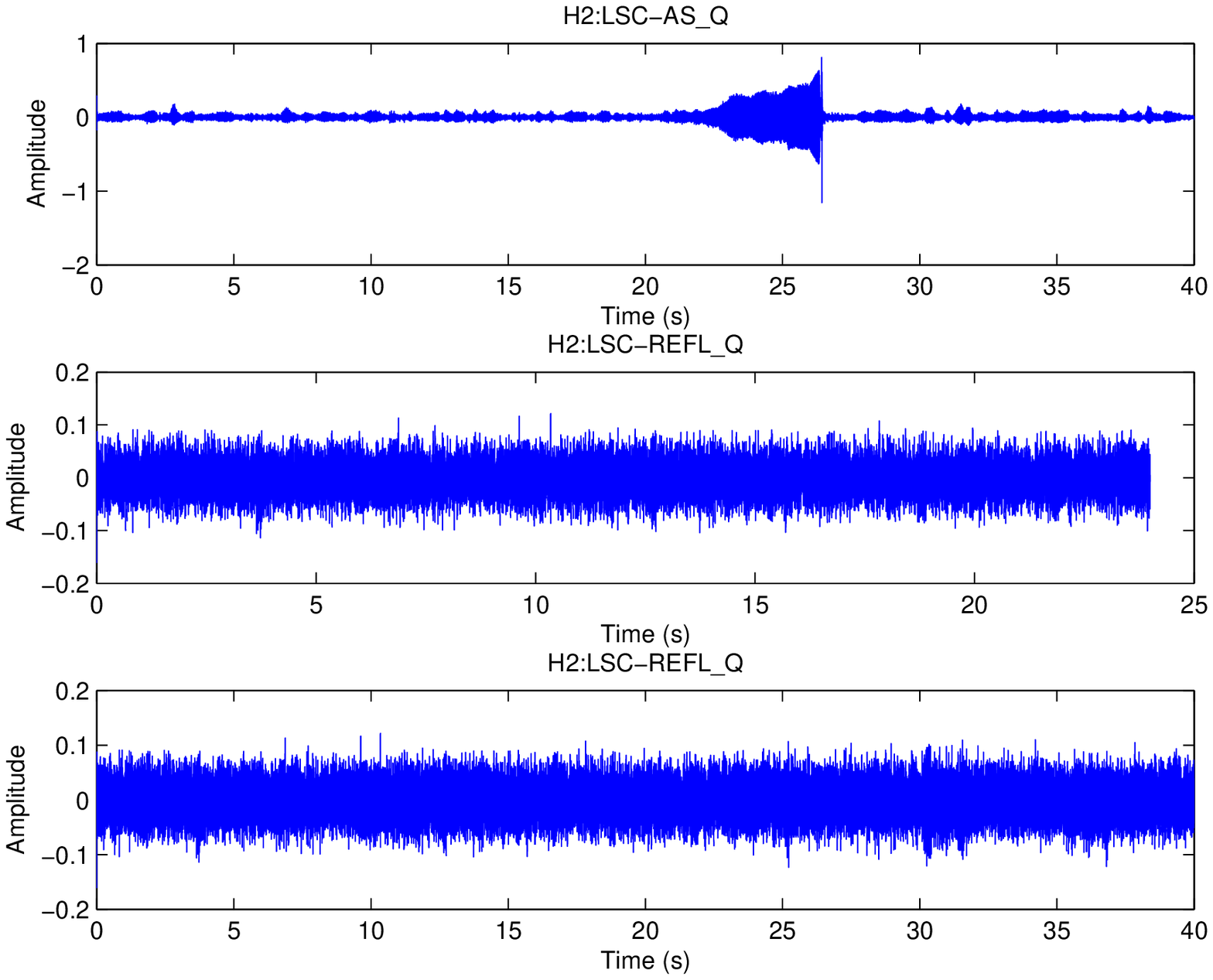}
  \end{center}
\caption{An example of an inspiral hardware injection in H2, which can be seen in channel H2:LSC-AS\_Q. The injection is not to be seen in either H2:LSC-REFL\_Q or H2:LSC-PRC\_CTRL. The plots for all three channels here were generated with data that was passed through a 100Hz to 300Hz band-pass filter.
} \label{fig5}
\end{figure}

The {\it safety} of a veto is of paramount importance. We would not want to throw away a real gravitational wave event. Vetoes that have
a small likelihood of eliminating gravitational wave signals are called {\it safe}. As such, we looked closely at candidate veto channels during hardware injections (simulated signals injected into the interferometer), and assured ourselves that the signals in the gravity wave output channel of the interferometer do not also appear in the veto channels. We visually examined H2:LSC-REFL\_Q and H2:LSC-PRC\_CTRL during 17 hardware injections, and both channels were seen to be safe. As an example, Fig.~\ref{fig5} displays a hardware injection in H2:LSC-AS\_Q, and along with the response of our two veto channels.

The veto performance of H2:LSC-REFL\_Q and H2:LSC-PRC\_CTRL are basically equivalent. They also tend to veto the same events, so using both together as a veto does not increase the efficiency. The LSC S3 burst event search for H2 will use H2:LSC-PRC\_CTRL as a veto~\cite{S3-burst-veto}. In an attempt to be consistent with the use of vetoes we are considering the choice of H2:LSC-PRC\_CTRL as the S3 H2 inspiral veto.  

\begin{figure}[tb]
  \begin{center}
    \includegraphics[width=5.0in,angle=0]{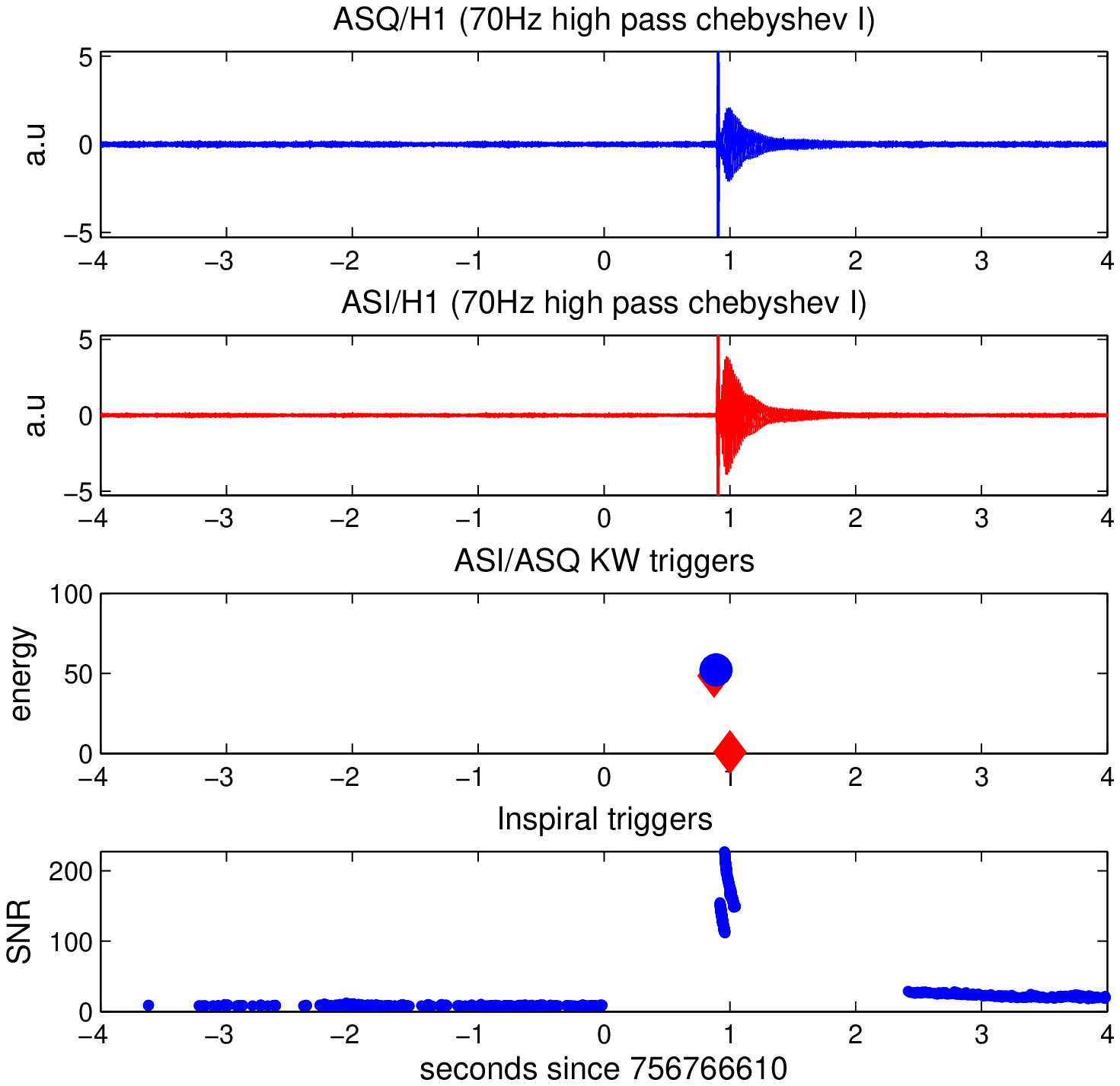}
  \end{center}
\caption{An example of simultaneous glitches seen in H1:LSC-AS\_Q and H1:LSC-AS\_I. Also displayed are the triggers from the KleineWelle wavelet based glitch finding program~\cite{S3-burst-veto}; circles for AS\_Q and diamonds for AS\_I. At the bottom is the distribution of inspiral triggers produced from this glitch.
} \label{fig6}
\end{figure}

With H1 inspiral events we are considering the use of 
channel H1:LSC-AS\_I as a veto.
While H1:LSC-ASQ\_Q is linearly related to
the gravitational wave strain, H1:LSC-AS\_I contains
information on optical imbalances and mirror misalignments.
Investigations of the loudest inspiral triggers in the S3 H1
playground revealed significant coincident
disturbances in the two channels resulting from the
radio-frequency demodulation of the interferometer dark port photocurrent.  
A typical glitch is shown in Fig.~\ref{fig6}.  
Using triggers produced by the KW wavelet analysis \cite{S3-burst-veto,KW} 
to find disturbances in the H1:LSC-AS\_I channel, 
we developed a veto strategy that
successfully removed inspiral triggers with a high SNR.

Once again, scrutiny must follow any proposed veto strategy in order to avoid
accidentally vetoing gravitational waves. 
Veto safety is most important for the AS\_I channel, since phase error in the
demodulation of AS\_Q and AS\_I results in a mixing of those
channels.  H1:LSC-AS\_I was monitored during hardware injections. 
To safely use H1:LSC-AS\_I as a veto we put restrictions on
the relative amplitudes of glitches in both channels (AS\_I and AS\_Q) 
compared to the
relative amplitude of the simulated gravitational wave signals. We
compared the energy of the KW triggers in AS\_I and AS\_Q 
during the injection
times to establish a safe ratio between glitches and "real signals" with
those channels.  We also imposed a minimum energy threshold on the AS\_I KW
triggers to reduce the deadtime incurred when applying this strategy,
without reducing the efficiency significantly.

\begin{figure}[tb]
  \begin{center}
    \includegraphics[width=5.0in,angle=0]{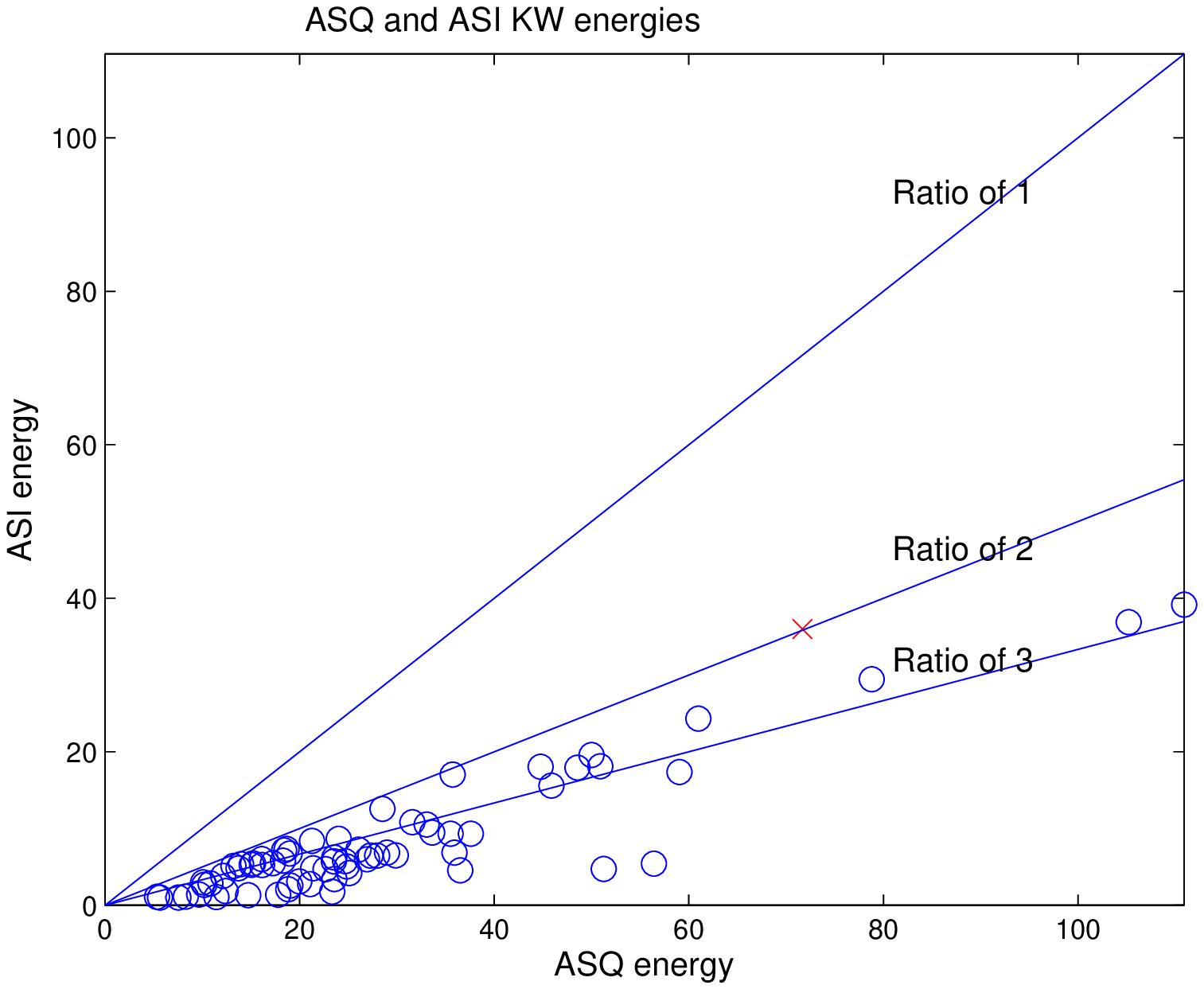}
  \end{center}
\caption{The KleineWelle glitch finding program was run on sections of data containing hardware inspiral injections. Plotted is the distribution of KleineWelle trigger energies for H1:LSC-AS\_Q and H1:LSC-AS\_I data. The one trigger marked with an $\times$ has the smallest ratio ($\approx 2$) between AS\_Q and AS\_I
} \label{fig7}
\end{figure}

Our criteria for creating a veto was the following: we required AS\_I 
KW triggers to have AS\_Q KW trigger counterparts;  
we required a minimum energy threshold in the AS\_I
KW triggers, and we required a safe ratio of energies in AS\_I and AS\_Q. 
The KW triggers that satisfied these conditions were considered veto
candidates. In Fig.~\ref{fig7} we plot the KW energy 
for AS\_I and AS\_Q for 60 hardware injections of binary inspiral signals.
We looked for coincidences between our veto candidates and
inspiral triggers and found that a large percentage
were used successfully to veto inspiral triggers. For example, with a 
window from -1s to +8s about the KW H1:LSC-AS\_I trigger, and vetoing inspiral
triggers when the ratio of KW energy for H1:LSC-AS\_Q to KW energy 
for H1:LSC-AS\_I  is less than 2.0, the veto efficiency is 72\%, with a 
resulting deadtime of only 0.33\%. There was one hardware injection (see the $\times$ in Fig.~\ref{fig7}) event with a AS\_I to AS\_Q KW energy ratio of 2, so a safe veto should have a threshold that is smaller.

Though we looked carefully, no effective and safe veto was found for S3 inspiral events at L1. L1:LSC-AS\_I was studied closely, as numerous  glitches in the data were observed in coincidence with inspiral triggers. However, the veto could not be demonstrated to be completely safe; the ratio of KW energies for AS\_Q and AS\_I glitches was comparable to the ratio for hardware injections.

\section{Summary} 
\label{Disc}
The LSC is actively searching for binary inspiral signals in its S3 data. Many interferometer control channels and environmental monitors were inspected during the times of inspiral triggers (within the playground period of S3). We have identified a number of effective data quality flags. In addition, we are applying vetoes to the inspiral triggers from H1 and H2. We have been able to identify classes of glitches that are caused by imperfections in interferometer performance. Through the examination of the veto channels during inspiral hardware injections we are able to show that these are safe vetoes, and a real gravitational wave event will not be excluded by them.
  
For the inspiral analysis of the S3 LIGO data we are also considering the implementation of a signal based veto. This veto is based on the work of ~\cite{Shawhan}, and monitors the evolution of the SNR with time through the event. Further details on the signal based veto will be given in a forthcoming publication.

\verb''\ack
LIGO Laboratory and the LIGO Scientific Collaboration gratefully acknowledge
the support of the United States National Science Foundation for the construction and operation of the LIGO Laboratory and for the support of this research. The author's work is supported by National Science Foundation grant PHY-0244357.

\end{document}